\documentclass[%
preprint,
superscriptaddress,
nofootinbib,
nobibnotes,
 amsmath,amssymb,
 aip,
]{revtex4-1}

\usepackage{siunitx}
\usepackage{graphicx}
\usepackage{dcolumn}
\usepackage{bm}
\usepackage{epstopdf}
\usepackage{bbm}
\usepackage{amsmath}
\usepackage[hidelinks]{hyperref}
\usepackage[utf8]{inputenc}
\usepackage[T1]{fontenc}
\usepackage{mathptmx}

\usepackage{float}
\usepackage{color}

\begin{document}

\preprint{AIP/123-QED}

\title{Sublattice symmetry breaking and ultra low energy excitations in Graphene-on-hBN Heterostructures}

\author{U. R. Singh}
\affiliation{%
	Center for Hybrid Nanostructures (CHyN), University of Hamburg, Luruper Chaussee 149, 22607 Hamburg, Germany
}
\author{M. Prada}%
 \email{mprada@physnet.uni-hamburg.de}.
\affiliation{%
	I. Institute for Theoretical Physics, University of Hamburg, Jungiusstrasse 9-11, 20355 Hamburg, Germany
}
\author{V. Strenzke}%
\affiliation{%
	Center for Hybrid Nanostructures (CHyN), University of Hamburg, Luruper Chaussee 149, 22607 Hamburg, Germany
}
\author{B. Bosnjak}%
\affiliation{%
	Center for Hybrid Nanostructures (CHyN), University of Hamburg, Luruper Chaussee 149, 22607 Hamburg, Germany
}
\author{T. Schmirander}%
\affiliation{%
	I. Institute for Theoretical Physics, University of Hamburg, Jungiusstrasse 9-11, 20355 Hamburg, Germany
}
\author{L. Tiemann}%
\affiliation{%
	Center for Hybrid Nanostructures (CHyN), University of Hamburg, Luruper Chaussee 149, 22607 Hamburg, Germany
}
\author{R.H. Blick} \altaffiliation[ Also at ]{Materials Science and Engineering,
University of Wisconsin-Madison, 1550 University Avenue, Madison, Wisconsin 53706, USA }
\affiliation{%
	Center for Hybrid Nanostructures (CHyN), University of Hamburg, Luruper Chaussee 149, 22607 Hamburg, Germany
}
\date{\today}
\begin{abstract}
The low-lying states of graphene contain exciting topological properties that depend on the interplay of different symmetry breaking terms. The corresponding energy gaps remained unexplored until recently, owing to the low energy scale of the terms involved (few tens of \si{\micro}eV). These low energy terms include sublattice splitting, the Rashba and the intrinsic spin-orbit coupling, whose balance determines the topological properties. In this work, we unravel the contributions arising from the sublattice and the intrinsic spin orbit splitting in graphene on hexagonal boron-nitride. Employing resistively-detected electron spin resonance, we measure a sublattice splitting of the order of 20 \si{\micro}eV, and confirm an intrinsic spin orbit coupling of approximately 45 \si{\micro}eV. The dominance of the latter suggests a topologically non-trivial state, involving fascinating properties. Electron spin resonance is a promising route towards unveiling the intriguing band structure at low energy scales.
\end{abstract}

\keywords{graphene, hexagonal boron nitride, heterostructures, microwave spectroscopy, electron spin resonance, spin-orbit interaction.}
\maketitle

The most striking electronic properties of graphene are linked to the bipartite nature of the honeycomb lattice, consisting of two interpenetrating triangular sublattices. In idealized graphene with sublattice degeneracy,  the electrons become chiral in the continuum limit \cite{neto}. The electrons can then be described by a Dirac Hamiltonian, which accounts for the absence of back scattering \cite{kastnelson}. Atomic – or more specifically – intrinsic spin orbit coupling (SOC) breaks spin rotational symmetry, preserving chirality and opening a gap in the spectrum \cite{kane1,kane2,sichau}. This gap transforms the system into a topological insulator that gives rise to the spin-Hall effect. Sublattice symmetry breaking terms can preserve the topological properties as long as the splitting is compensated by the intrinsic SOC gap. Resolving the low-lying energetic bands in graphene is thus crucial, in order to unveil the rich physics of Dirac charge carriers. We employ resistively-detected electron spin resonance (RD-ESR) and access these low-lying bands.

Sublattice symmetry breaking in van der Waals (vdW) materials can arise due to coupling to the substrate \cite{Song-2013, Jung-2015}. An example for this is the encapsulation of the charge carrying material graphene with insulators, such as hexagonal boron nitride (hBN) \cite{Dean-2010, Yankowitz-2013, Hunt-2013, Wang-2015}. Having a small lattice mismatch of about 1.8$\%$ between the two hexagonal lattices \cite{Song-2013, Yankowitz-2013}, the symmetry breaking favors energetically one of the two sublattices A and B in graphene on hBN (GohBN) \cite{Song-2013, Jung-2015, Hunt-2013, Amet-2013}. This can lead to the modification of the band structure including the opening of a gap, which remains an experimentally unexplored realm \cite{Song-2013, Jung-2015, zollner}.

In this paper, we address the phenomenon of spin splitting of sublattices A and B in GohBN, both theoretically and experimentally. We determine the size of splitting gap using resistively-detected electron spin resonance at low temperatures. Our findings reveal the coexistence of intrinsic spin-orbit coupling and the spin splitting of sublattices with very small splitting gap values.

In the Dirac model, widely used to describe Dirac carriers in graphene \cite{neto,kastnelsonB}, the notion of sublattice spin is introduced, where first neighbor hopping is identified with a spin-flip operator. We describe the Hamiltonian using a minimal model in the bi-spinor basis spanned by spin and sublattice spin,
$\{\uparrow, \downarrow\}\otimes\{A,B\}$,
\[
\hat H = \hbar v_F \mathbb{I}\otimes\left(
\tau \hat \sigma_x k_x + \hat \sigma_y k_y \right) +
\frac{1}{2}\Delta_I\tau \hat s_z \otimes\sigma_z
+ \frac{1}{2}\Delta_\gamma\hat s_z \otimes\mathbb{I} ,
\]
where $\Delta_I$ is the intrinsic SOC gap, $\Delta_\gamma$ the sublattice splitting, $\tau$ the valley index and $(k_x, k_y)$ the small vector near the DP \cite{neto}. Without losing generality, we consider the case
$\Delta_I > \Delta_\gamma$. In the absence of Rashba SOC, the conduction band (CB) eigenstates near the K point ($\tau$= 1, $k_x = k_y$  = 0) are $\{\uparrow, \rm A\}$ and $\{\downarrow, \rm B\}$ with energies
$\Delta_I + \Delta_\gamma$  and $\Delta_I - \Delta_\gamma$  , respectively, whereas the CB eigenstates at K$^\prime$ are $\{\downarrow, \rm A\}$ and $\{\uparrow, \rm B\}$, with energies $\Delta_I + \Delta_\gamma$  and $\Delta_I - \Delta_\gamma$.

\begin{figure}[!hbt]
\centering
\includegraphics[angle=0, width=.99\linewidth]{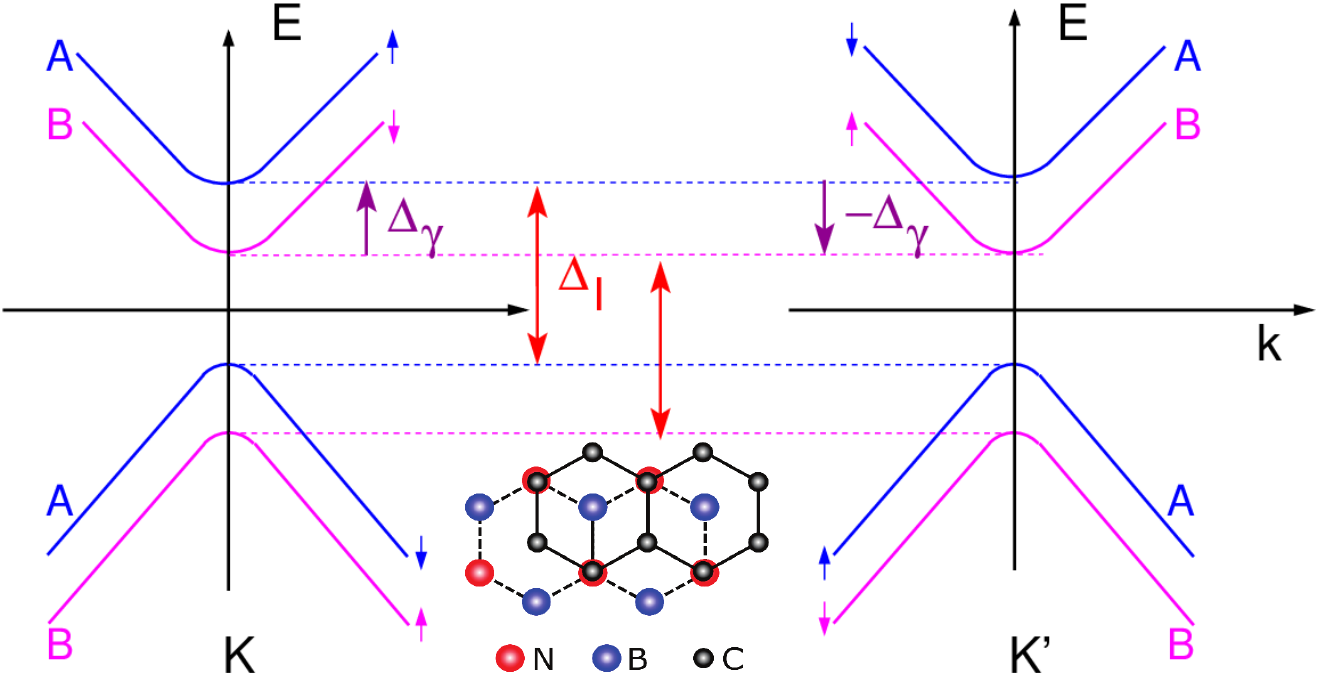}\\
\caption{
Schematic band structure of the bulk bands near the DPs at $B =0$ in the topological insulator condition: $\Delta_I>\Delta_\gamma$. The $s_z$ eigenvalue of the bands is marked with an arrow, whereas the colors indicate sublattice degree of freedom: Blue for sublattice A and magenta for sublattice B. Inset: a case for boron ($B$) atoms of hBN situated at the centre of the hexagonal lattice of graphene.
}
\label{fig1}
\end{figure}

Fig. \ref{fig1}  shows schematically the band structure at both Dirac points (DPs), which are related by time inversion (and thus spin). The bands are colored according to the sublattice (blue for sublattice A, and magenta for sublattice B), and the spin eigenvalue $s_z$ is denoted with an arrow. As a result of the chirality of graphene, the sublattice gap in neighboring DPs has opposite sign. As a consequence, a magnetic field splits further the CB levels at one DP and reduces splitting at the other, that is, the splitting between CB spin up and down bands at either DP would be,
\begin{equation}
\label{eq1}
\Delta_\varepsilon^\pm =  g \mu_{\rm B} B\pm   \Delta_\gamma.
\end{equation}

Here, $\mu_{\rm B}$ is the Bohr magneton and $g$ is the $g$-factor, which is close to 1.95 in these experiments \cite{sichau, lyon}. Our goal is to address these opposite-spin bands by employing RD-ESR, a method that had previously verified the existence of the intrinsic band gap in graphene \cite{sichau, mani}. This spin-sensitive probing technique allows us to resolve the energetic distance between spin bands, as the microwave excitation couples opposite spins \cite{slichter}. We can detect the response to the resonant absorption of microwaves as changes in the resistivity in a Hall bar structure.

\begin{figure}[!h]
\centering
\includegraphics[angle=0, width=.9\linewidth]{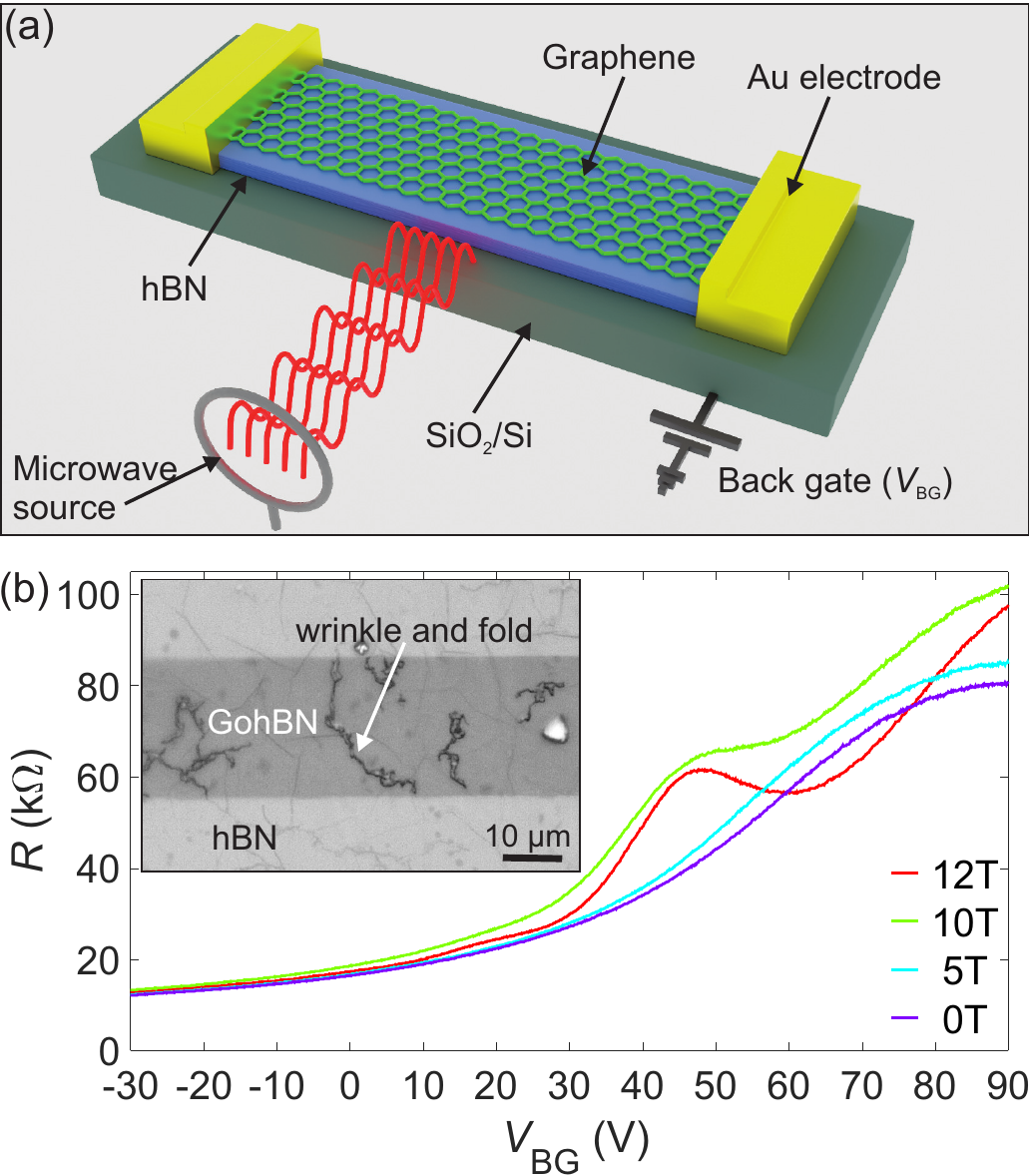}
\caption{(a) Schematic illustration of our 200 \si{\micro}m $\times$ 22 \si{\micro}m GohBN Hall bar device and the loop antenna for microwave irradiation. (c) The apex of $R$ vs $V_{BG}$ at 90 V for $B$ = 0 T marks the charge neutrality point. Measurements at finite magnetic fields show the onset of Landau quantization. Inset: Optical image of GohBN Hall bar showing the surface of CVD graphene as a darker region.
}
\label{fig2}
\end{figure}

The 200 \si{\micro}m $\times$ 22 \si{\micro}m GohBN-Hall bars were fabricated using chemical vapor deposition (CVD)-graphene and CVD-hBN on a highly $p$-doped Silicon-on-Insulator (SOI) substrate with 285 nm top layer of SiO$_2$. The CVD-hBN obtained on copper foil\cite{grap-supmark} was removed in iron nitrate solution, and after its transfer to the SOI, annealing was performed at $\sim$180 $^\circ$C to ensure adhesion to the substrate \cite{Shautsova-2016, Fazio-2019}. We fabricate the Hall bar structure by transferring graphene onto the hBN and using optical lithography and ion etching, as described in detail elsewhere \cite{lyon2}. The resulting GohBN sample was thermally annealed at 350$^\circ$C for two hours to remove chemical residues remaining on the surface and in the vicinity of interfaces \cite{Yankowitz-2013, Kim-2015}. Metallic contacts were fabricated by depositing Ti/Au (7 nm/70 nm) via photolithography and physical vapor deposition.

\begin{figure}[!h]
\centering
\includegraphics[angle=0, width=.9\linewidth]{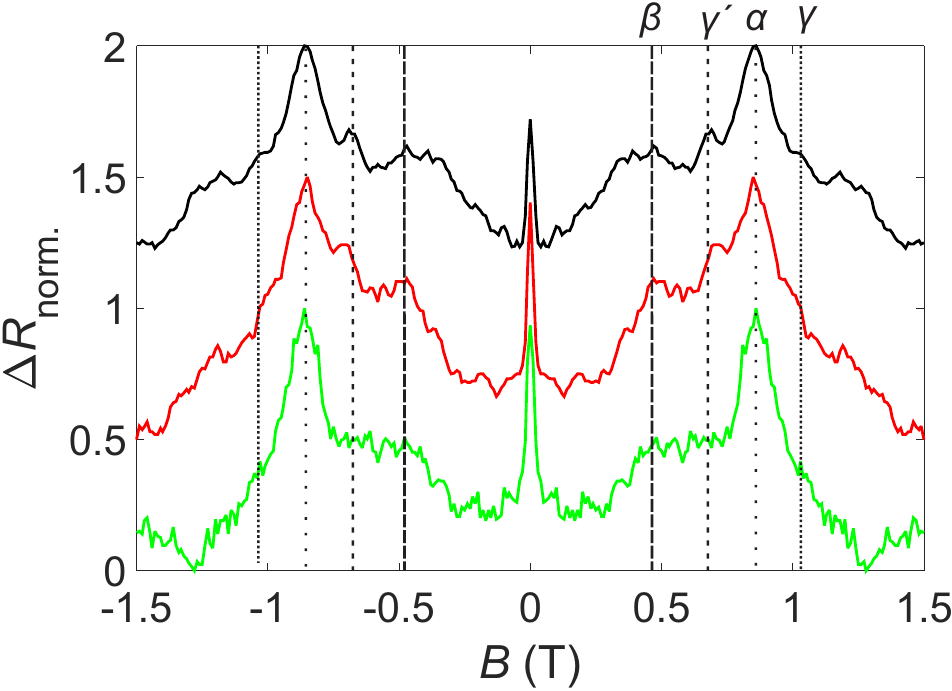}\\
\caption{
Normalized curves of $\Delta R$ versus $B$ at the exemplary frequency $f$ of 23 GHz (21 dBm) plotted for $V_{\rm BG} $ = 0 V (green solid line), 50 V (red solid line), and 85 V (black solid line). Resonances between spin-split bands are labeled $\alpha$, $\beta$, $\gamma$ and $\gamma^\prime$. The vertical lines are a guide to the eye.}
\label{fig3}
\end{figure}

To be able to resolve low energy excitations in the energy spectrum of graphene on hBN, we cooled the sample down to $\sim 1.3$ K in vacuum and used a low-frequency lock-in technique for the detection of the magnetoresistance and ESR under perpendicular magnetic fields ($B$). A back gate voltage ($V_{\mathrm{BG}}$) was applied to the substrate to vary the carrier concentration in the graphene layer and to shift the Fermi level closer to the DP. The microwaves are generated by a loop antenna placed next to the sample as illustrated in Fig. \ref{fig2}(a). Our graphene device is subject to unintentional doping; a back gate voltage of 90 V is required to access the charge neutrality point (CNP), seen as an apex in the two-terminal resistance [Fig. \ref{fig2}(b)]. A perpendicular $B$ allows us to observe Landau quantization presented in the same figure. Further, we have extracted the carrier density $n$ from the Hall measurement and obtained the carrier mobility ($\mu$), which is of the order of 1200 \si{\square\cm\per\volt\per\second}. The poor mobility in our device can be limited by scattering on grain boundaries, wrinkles and folds that are inherent to large scale van der Waals materials synthesized by CVD and the transfer process \cite{Kim-2015, Petrone-2012, Zhu-2012}. The optical microcopic image in the inset of Fig. \ref{fig2}(b) visualizes some of these defects in our device.

\begin{figure*}[!t]
\centering
\includegraphics[angle=0, width=.95\linewidth]{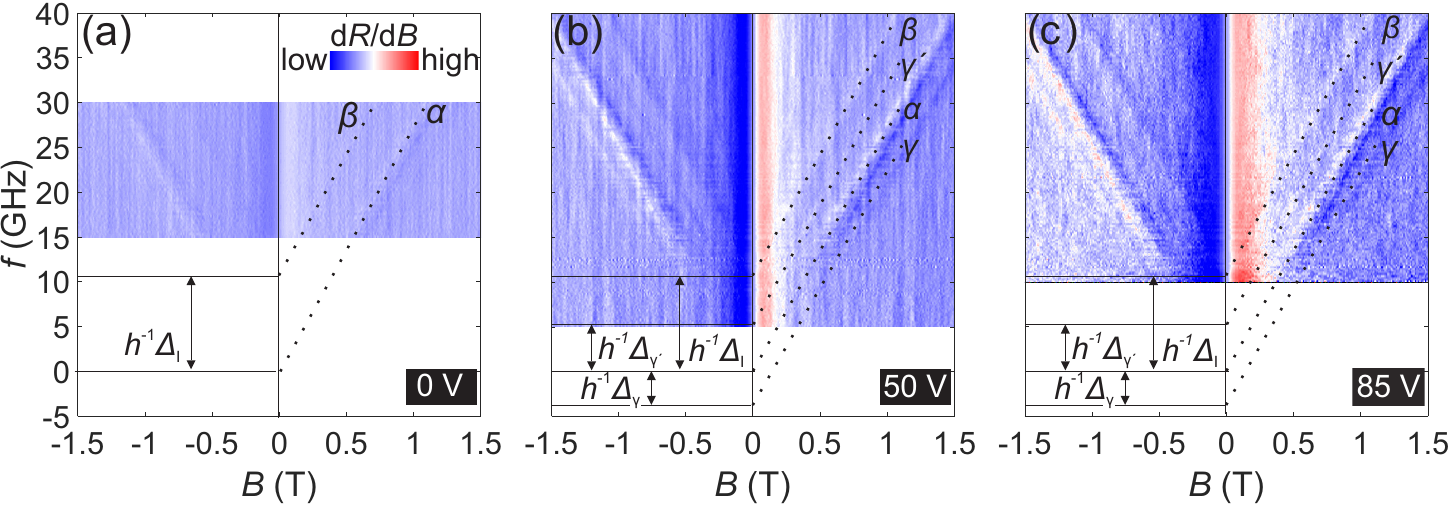}\\
\caption{
$\mathrm{d}R/\mathrm{d}B$ as a function of $B$ and $f$ for $V_{\rm BG}$ = 0 V (a), 50 V (b), and 85 V (c). $\alpha$ extrapolates to the origin in all measurements. $\beta$ intensifies notably closer to the CNP, and the intercept of the $\beta$ line with the $f$-axis represents intrinsic SOC. As a result of sublattice splitting, $\gamma$ and $\gamma^\prime$ emerge out of the resistive background close to the CNP.
}
\label{fig4}
\end{figure*}

We can identify low energy interband excitations between spin split bands by probing the longitudinal sample resistance $R_{xx}$ as a function of $B$ in the absence ($R_{\rm xx}^{\rm dark}$) and in the presence ($R_{\rm xx}^{\rm RF}$) of microwave radiation of constant frequency. The microwave radiation thermally activates carriers, thus reducing the overall resistance as new channels for transport become available \cite{datta}. Whenever the microwave energy $h\cdot f$ matches the energetic difference between two opposite spin polarized bands, an additional \textbf{resonant} interband excitation leads to a distinct peak in the photo-induced differential resistance $\Delta R = R^\mathrm{dark}_\mathrm{xx} -R^\mathrm{RF}_\mathrm{xx}$.

Figure \ref{fig3} shows the resulting (normalized) resistance $\Delta R_{norm}$ as a function of $B$ for an exemplary frequency $f$ of 23 GHz, measured with $V_{BG}$ = 0 V (i.e., energetically distant to the CNP), 50 V, and 85 V (i.e., energetically close to the CNP). $\Delta R_{norm}$ was obtained by normalizing each $\Delta R$ to the range 0 to 1. Resonant features are marked by dotted lines, and we  introduce the labeling convention $\alpha$, $\beta$, $\gamma$, and $\gamma^\prime$ to distinguish these resonances. We note that at this frequency not all resonances are pronounced.

We have repeated the measurements at various constant frequencies. Figures \ref{fig4}(a), (b) and (c) show the derivative of the two-terminal resistance ($\mathrm{d}R/\mathrm{d}B$) in the $f$-$B$ plane for $V_{BG}$ = 0 V, 50 V, and 85 V. For 0 V in Fig. \ref{fig4}(a), at high carrier concentrations and energetically distant from the CNP, we only observe the linear dispersions of the $\alpha$ and $\beta$ resonances. The linear extrapolation of $\alpha$ intercepts with the origin, suggesting that this line represents the ordinary Zeeman splitting. $\beta$ intercepts  with the $f$-axis at around 10 GHz and represents the intrinsic SOC gap \cite{sichau}. For $V_{BG}$ = 50 V and 85 V in Fig. \ref{fig4}(b) and \ref{fig4}(c), i.e., at lower carrier concentrations and energetically closer to the CNP, $\alpha$ and $\beta$ are accompanied by the satellite resonances $\gamma$ and $\gamma^\prime$ which intercept with the $f$-axis at h$^{-1}\Delta_\gamma$ < 0 and h$^{-1}\Delta_{\gamma^{\prime}}$ > 0, respectively.

\begin{figure}[!h]
\centering
\includegraphics[angle=0, width=.9\linewidth]{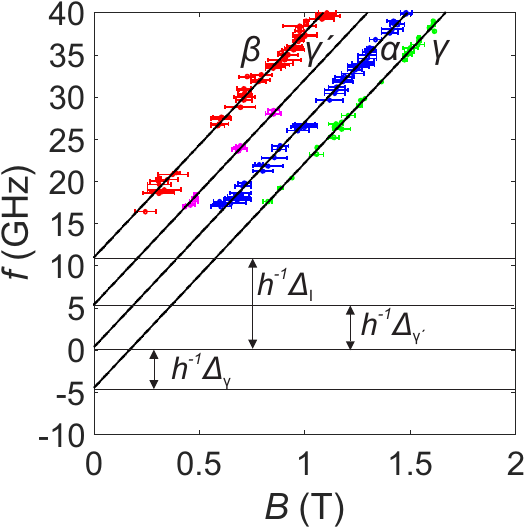}
\caption{Frequency vs absolute magnetic field, where resonances occurred, obtained by determining the zero-crossings of $\mathrm{d}R/\mathrm{d}B$ from data in Fig. \ref{fig4}(c). The length of the error bars is estimated by taking the difference between the zero-crossing and maximum and minimum in the $\mathrm{d}R/\mathrm{d}B$ curves. All black solid lines represent linear fits. The $\alpha$ fit yields a slope of $\Delta f/\Delta B$ = (26.68 $\pm$ 0.48) GHzT$^{-1}$, or a $g$-factor of 1.90$ \pm$ 0.04, respectively. This $g$-factor was used to fit $\beta$, $\gamma$ and $\gamma^{\prime}$ and obtain the intercepts with the $f$-axis of h$^{-1}\Delta_{I}$ $\approx$ (10.94 $\pm$ 0.19) GHz, h$^{-1}\Delta_{\gamma^{\prime}}$ $\approx$ (5.35 $\pm$ 0.15) GHz, and h$^{-1}\Delta_{\gamma}$ $\approx$ (-4.52 $\pm$ 0.15) GHz.
}
\label{fig5}
\end{figure}

The emergence of the $\gamma$ and $\gamma^\prime$ lines can be reconciled with a sublattice splitting according to equation (\ref{eq1}) that is induced by the interaction with the hBN. The microwaves address these spin- and sublattice split bands at low energies \cite{zollner, Cummings-2019} and trigger additional interband excitations. We stress that extrinsic (i.e., Rashba) SOC contributions cannot account for the existence of $\gamma$ and $\gamma^\prime$. A hypothetical Rashba contribution would linearly increase with the electric field\cite{rashba} from the back gate, and would \textbf{not} be symmetric with respect to $\alpha$. According to Ref. \onlinecite{zollner}, at least 5 Volts$/$nm are needed to observe extrinsic spin-orbit coupling effects. The dielectric layer in our sample, consisting of hBN and almost 300 nm of Si oxide, is too thick to generate such a field strength.

We can obtain the spin and sublattice splitting energies by carefully analyzing the resonance ocurrences presented in Fig. \ref{fig4}(c). Figure \ref{fig5} shows the results of this analysis as frequency $f$ versus the magnetic field occurrence of the resonance. A linear fit of the data points from $\alpha$ yield the slope $\Delta f/\Delta B$ = (26.68 $\pm$ 0.48) GHz$\cdot$T$^{-1}$ and a $g$-factor of 1.90 $\pm$ 0.04, which is consistent with previous reports on single and multilayer graphene on SiO$_2$ and SiC substrates\cite{sichau, mani}. We then use 1.90 $\pm$ 0.04 to linearily fit $\beta$, $\gamma$ and $\gamma^{\prime}$ and obtain the individual intercepts $h^{-1}\Delta$ with the frequency axis. The intercept times Planck's constant $h$ yields the transition energies of $\Delta_{I} = (45.24\pm 0.79)~$\si{\micro}eV for $\beta$, $\Delta_{\gamma^{\prime}} = (22.13\pm 0.62)~$\si{\micro}eV for $\gamma^{\prime}$ and $\Delta_\gamma = (18.69\pm 0.62)~$\si{\micro}eV for $\gamma$.

We can now compare our results to anticipated interband transitions in graphene that are subject to spin \textbf{and} sublattice splitting, as illustrated in Fig. \ref{fig1}. The $\alpha$ and $\beta$ resonances represent excitations between spin split bands and across the intrinsic band gap, as previously reported \cite{sichau}. The $\gamma$ and $\gamma^{\prime}$ spin resonances, however, can only occur when additionally to the lifting of the (electron) spin degeneracy also the degeneracy of sublattices A and B is lifted. 

We point out that the combination of monolayer graphene and boron nitride results in strong interactions between the two materials and in the emergence of Moir\'{e} superlattices \cite{Yankowitz-2013}. On a  microscopic scale, the periodic potential created due to graphene superlattices can lead to a modification of the band structure at a low energy scale, giving rise to an asymmetry between the sublattices A and B and thus sublattice symmetry breaking \cite{Song-2013, Jung-2015, Hunt-2013, Amet-2013, Wang-2015}. The locations of $B$ and $N$ under carbon atoms of graphene play an important role in defining the sublattice splitting \cite{zollner, Cummings-2019}. Van der Waals materials synthesized by CVD are comprised of many grains and the randomized locations of the atoms are expected to result in some averaged value for the sublattice splitting. In our device, we find the energy scale of the sublattice splitting to be of the order of 20 \si{\micro}eV.

For completion, we want to note that we also extracted the spin diffusion length\cite{Kamalakar-2015, Gurram-2018} $\lambda_{s}=\sqrt{D\tau_{s}}$ ($D$: diffusion constant), from the $\alpha$ resonances in Fig. \ref{fig4}(c). The spin lifetime $\tau_{s}= \hbar\cdot(2h\frac{\Delta f}{\Delta B})^{-1}$, which is proportional to the resonance peak width ($\Delta B$), yields a $\lambda_{s}$ ranging between 350 nm to 550 nm. 

In conclusion, we demonstrated that the interaction between graphene and hBN lifts the degeneracy of sublattices A and B. The sublattice splitting energy, which in our device is found to be of the order of 20 \si{\micro}eV, is minuscule and would be dwarfed by the energetic resolution of most experimental techniques. Our RD-ESR experiments, however, we can address the individual spin split bands, giving us access to the realm of ultra low energy scales and complex band structures that are subject to competing interactions.

We acknowledge support by the Partnership for Innovation, Education, and Research (PIER). We also thank the Excellence Cluster Center for Ultrafast Imaging (CUI) of the Deutsche Forschungsgemeinschaft (DFG) for support under contract number EXC-1074. We are grateful to Jann Harberts for rendering the image shown in Fig. 2(a). All measurements in this work were performed with \textsl{nanomeas}.

\bibliographystyle{plain}
\bibliography{main}

\end{document}